\documentstyle[12pt]{article}

\begin{document}
\title{
  \vspace*{-4ex} \hfill {\large UT-738 \ \ \ } \\
  \vspace{5ex}
  Nonabelian Duality and Higgs Multiplets \\
  in Supersymmetric Grand Unified Theories
  }
\author{T. Hotta, Izawa K.-I. and T. Yanagida\\
  \\  Department of Physics, University of Tokyo \\
  Bunkyo-ku, Tokyo 113, Japan}
\date{\today}
\maketitle
\setlength{\baselineskip}{3.6ex}
\begin{abstract}

We consider strongly interacting supersymmetric gauge theories which
break dynamically the GUT symmetry and produce the light Higgs
doublets naturally.
Two models we proposed in the previous articles are reanalyzed as two
phases of one theory and are shown to have desired features.
Furthermore, employing nonabelian duality proposed recently by
Seiberg, we study the dual theory of the above one and show that the
low-energy physics of the original and dual models are the same as
expected.
We note that the Higgs multiplets in the original model are regarded
as composite states of the elementary hyperquarks in its dual theory.
Theories with other hypercolors and similar matter contents are also
analyzed in the same way.

\end{abstract}

\newpage

\section{Introduction}

The supersymmetric grand unified theory (SUSY-GUT) \cite{susygut} is
one of the promising candidates for the physics beyond the standard
model.
In fact, the recent high-precision measurements on the standard-model
parameters such as the Weinberg angle agree with some of its
predictions \cite{preciceexp}.
In spite of the remarkable success, there is a fatal fault in the
SUSY-GUT: a fine-tuning problem.
Since the GUT scale, typically $10^{16}$ GeV, is extremely high
compared with the weak scale $\sim 10^2$ GeV, we have to adjust
parameters in the GUT accurately in order to have a Higgs doublet in
the standard model.
Although a number of attempts have been made to solve this serious
problem, there was no convincing model to explain the origin of the
light Higgs doublet.

In recent papers \cite{our1, our2, our3} we have proposed SUSY gauge
theories whose interactions are strong at the GUT scale causing
dynamical breaking of the GUT symmetry.
These models also provide mechanisms which produce the light Higgs
doublet naturally.

The main purpose of this paper is to examine the strongly interacting
SUSY gauge theories more thoroughly.
In addition to the method used to find quantum vacua in
Ref.\cite{our2, our3}, we employ nonabelian duality which has been
proposed recently by Seiberg \cite{seiberg1} as a powerful tool to
investigate nonabelian gauge theories.
Since the nonabelian duality states that SU($N_c$) and SU($N_f - N_c$) 
gauge theories with the common $N_f$ flavors have the same low-energy
behavior, especially on the vacuum structure, we may reduce the number
of theories to study and also check consistency of the results using
both theories.

In section \ref{sec:original} we review the results of our two models
in Ref.\cite{our2, our3} and show that these models are regarded as
two phases of one theory.
In section \ref{sec:dual} we consider the dual theory of the model in
section \ref{sec:original} and show that the low-energy physics of the 
dual model is the same as that of the original one.
We, however, stress that short-distance structures of the original and 
dual models are different from each other and hence these models
represent different physics above the GUT scale.
We also note that the Higgs multiplets in the original model are
composite states of the elementary hyperquarks in the dual theory.
In section \ref{sec:hypercolor} we extend our analysis to theories
with other hypercolors and similar matter contents.
Section \ref{sec:conclusion} is devoted to our conclusions.
We also comment on some extensions and modifications of the models.

\section{The original model}
\label{sec:original}

We review the models studied in Ref.\cite{our2, our3} in which light
Higgs doublets are generated dynamically.
We analyze these two models in a unified manner treating them as
different phases of a single theory.

The model is based on a supersymmetric hypercolor SU(3)$_H$ gauge
theory with six flavors of hyperquark chiral superfields $Q_\alpha^A$
and $\bar{Q}^\alpha_A$ $(\alpha = 1, \cdots ,3 ; A = 1, \cdots, 6)$ in
the fundamental representations {\bf 3} and $\bf 3^*$ of SU(3)$_H$,
respectively.
The first five $Q_\alpha^I$ and $\bar{Q}^\alpha_I$ $(I = 1 , \cdots,
5)$ transform as $\bf 5^*$ and {\bf 5} under the GUT gauge group
SU(5)$_{GUT}$, respectively, while the last $Q_\alpha^6$ and
$\bar{Q}^\alpha_6$ are singlets of SU(5)$_{GUT}$.
We also introduce three kinds of SU(3)$_H$-singlet chiral superfields:
a pair of $H_I$ and $\bar{H}^I$, $\Sigma^I{}_{\! J}$ and $\Phi$ $(I, J
= 1, \cdots, 5)$ which are {\bf 5}, $\bf 5^*$, {\bf 24}+{\bf 1} and
{\bf 1} of SU$(5)_{GUT}$.

We impose a global U(1)$_A$ symmetry:
\begin{equation}
  \label{uonea}
  \begin{array}{ccl}
    Q_\alpha^I, \bar{Q}^\alpha_I & \rightarrow & Q_\alpha^I,
    \bar{Q}^\alpha_I , \\
    \noalign{\vskip 1ex}
    Q_\alpha^6, \bar{Q}^\alpha_6 & \rightarrow & e^{i \xi} Q_\alpha^6, 
        e^{i \xi} \bar{Q}^\alpha_6 , \\
    \noalign{\vskip 1ex}
    H_I, \bar{H}^I & \rightarrow & e^{- i \xi} H_I,
    e^{- i \xi} \bar{H}^I , \\
    \noalign{\vskip 0.5ex}
    \Sigma^I{}_{\! J} & \rightarrow & \Sigma^I{}_{\! J} , \\
    \noalign{\vskip 0.5ex}
    \Phi & \rightarrow & e^{-2 i \xi} \Phi , \\
    \noalign{\vskip 0.3ex}
    & & \hspace*{-4em} (I = 1, \cdots, 5) ,
  \end{array}
\end{equation}
to forbid such terms as $\bar{H}^I H_I$ and $\bar{Q}^\alpha_6
Q_\alpha^6$ in the superpotential.
Then, the superpotential is given by\,%
\footnote{We have chosen the normalization of the singlet field $Tr
  \Sigma$ so that the Yukawa term among $\Sigma^I{}_{\! J}$,
  $\bar{Q}^\alpha_I$ and $Q^I_\alpha$ is written with a single
  coupling constant $\lambda$ as shown in Eq.(\ref{orig.pot}).
  The effect of the rescaling of the field $Tr \Sigma$ appears in the
  K\"ahler potential, but it is irrelevant to the present analysis.}
\begin{equation}
  \begin{array}{rl}
    \label{orig.pot}
    W = & \lambda \Sigma^I{}_{\! J} \, \bar{Q}^\alpha_I \, Q^J_\alpha 
    + h H_I \bar{Q}^\alpha_6 Q^I_\alpha 
    + h' \bar{H}^I \bar{Q}^\alpha_I Q_\alpha^6
    + f \Phi \, \bar{Q}^\alpha_6 Q_\alpha^6 \\
    \noalign{\vskip 1ex}
    & \displaystyle + \frac{1}{2} \, m_\Sigma Tr(\Sigma^2)
    + \frac{1}{2} \, m'_\Sigma (Tr \Sigma)^2 - \mu_\Sigma Tr \Sigma
    \, . 
  \end{array}
\end{equation}
Here, we have omitted trilinear self-coupling terms of $\Sigma$ for
simplicity since they are irrelevant to the conclusion.
The global U(1)$_A$ has a strong SU(3)$_H$ anomaly and hence it is
broken by instanton effects at the quantum level.
However, as shown in Ref.\cite{our2} the broken global U(1)$_A$ even
plays a crucial role to protect a pair of massless Higgs doublets from 
having a mass.

Let us first consider a classical vacuum discussed in Ref.\cite{our2}:
\begin{equation}
  \label{classicalvacuum}
  \begin{array}{l}
    \langle Q_\alpha^A \rangle =
    \left(
      \begin{array}{ccc}
        0 & 0 & 0 \\
        0 & 0 & 0 \\
        v & 0 & 0 \\
        0 & v & 0 \\
        0 & 0 & v \\
        0 & 0 & 0
      \end{array}
    \right),\
    \langle \bar{Q}^\alpha_A \rangle =
    \left(
      \begin{array}{cccccc}
        0 & 0 & v & 0 & 0 & 0 \\
        0 & 0 & 0 & v & 0 & 0 \\
        0 & 0 & 0 & 0 & v & 0
      \end{array}
    \right), \\
    \noalign{\vskip 2ex}
    \displaystyle \langle \Sigma^I{}_{\! J} \rangle =
    \frac{\mu_\Sigma}{m_\Sigma + 2 m'_\Sigma}
    \left(
      \begin{array}{ccccc}
        1 & & & & \\
        & 1 & & & \\
        & & 0 & & \\
        & & & 0 & \\
        & & & & 0
      \end{array}
    \right) \, ,  
    \langle H_I \rangle = \langle \bar{H}^I \rangle = 0 \, ,
  \end{array}
\end{equation}
where
\begin{equation}
v = \sqrt{\frac{m_\Sigma \, \mu_\Sigma}{\lambda(m_\Sigma + 2
    m'_\Sigma)}} \, .
\end{equation}
Here, the vacuum-expectation value of $\Phi$ is undetermined since its 
potential is flat for $\langle Q_\alpha^6 \rangle = \langle
\bar{Q}^\alpha_6 \rangle = 0$.
In this classical vacuum the gauge group is broken down as
\begin{equation}
  SU(3)_H \times SU(5)_{GUT} \rightarrow SU(3)_C \times SU(2)_L \, .
\end{equation}
There is no unbroken U(1)$_Y$, and we introduce an extra U(1)$_H$
gauge symmetry in Ref.\cite{our1,our2} to have the standard-model
gauge group unbroken below the GUT scale $v$.%
\footnote{The GUT unification of three gauge coupling constants in the 
  standard model is realized in the strong coupling limit of U(1)$_H$
  \cite{our1}.}

Remarkable is that the missing partner mechanism \cite{missing} does
work very naturally in this classical vacuum \cite{our1}.
Namely, the color triplets $H_I$ and $\bar{H}^I$ $(I = 3, \cdots, 5)$
acquire the GUT-scale masses together with $Q_\alpha^6$ and
$\bar{Q}^\alpha_6$, respectively.
On the other hand the SU(2)$_L$-doublet $H_I$ and $\bar{H}^I$ $(I = 1,
2)$ remain massless, since they have no partners to form massive
chiral superfields with.

We now discuss quantum vacua where vacuum-expectation values of the
Higgs fields $\Sigma^I{}_{\! J}$, $H_I$ and $\bar{H}^I$ take forms
given in Eq.(\ref{classicalvacuum}).

In these vacua two hyperquarks $Q^I_\alpha$ and $\bar{Q}^\alpha_I$ $(I
= 1, 2)$ become massive.
The integration of the two massive hyperquarks leads to a low-energy
effective theory having the other four massless hyperquarks
$Q^I_\alpha$ and $\bar{Q}^\alpha_I$ $(I = 3, \cdots, 6)$.
We can then express the effective superpotential with meson
$M^i{}_{\! j}$, baryon $B_i$ and antibaryon $\bar{B}^i$ chiral
superfields
\begin{eqnarray}
  \label{defMeson}
  \begin{array}{lll}
    & & M^i{}_{\! j} \sim Q^i_\alpha \bar{Q}^\alpha_j \, , \\
    \noalign{\vskip 1ex}
    & & \displaystyle B_i \sim \epsilon^{\alpha \beta \gamma}
    \epsilon_{i j k l} Q^j_\alpha Q^k_\beta Q^l_\gamma \, , \\
    \noalign{\vskip 1ex}
    & & \displaystyle \bar{B}^i \sim \epsilon_{\alpha \beta \gamma}
    \epsilon^{i j k l} \bar{Q}_j^\alpha \bar{Q}_k^\beta
    \bar{Q}_l^\gamma \, , \\
    \noalign{\vskip 1ex}
  \end{array}
\end{eqnarray}
as follows \cite{our3}:
\begin{equation}
  \label{eff.pot}
  \begin{array}{rl}
    W_{eff} = & \Lambda^{-5} ( B_i M^i{}_{\! j} \bar{B}^j - \det
    M^i{}_{\! j} )
    + \lambda \Sigma^a{}_{\! b} M^b{}_{\! a} \\
    \noalign{\vskip 1ex}
    & + h H_a M^a{}_{\! 6} + h' \bar{H}^a M^6{}_{\! a}
    + f \Phi \, M^6{}_{\! 6} \\
    \noalign{\vskip 1ex}
    & \displaystyle - \frac{h h'}{m} (\bar{H}^1 H_1
    + \bar{H}^2 H_2) M^6{}_{\! 6} \\
    \noalign{\vskip 1ex}
    & \displaystyle + \frac{1}{2} \, m_\Sigma \Sigma^a{}_{\! b}
    \Sigma^b{}_{\! a}
    + \frac{1}{2} \, {m_\Sigma m'_\Sigma \over m_\Sigma + 2m'_\Sigma}
    ( \Sigma^a{}_{\! a} )^2
    - {m_\Sigma \mu_\Sigma \over m_\Sigma + 2m'_\Sigma}
    \Sigma^a{}_{\! a} \, ,
  \end{array}
\end{equation}
where $\Lambda$ denotes a dynamical scale of the low-energy SU(3)$_H$
interactions, $\displaystyle m = \frac{\lambda \mu_\Sigma}{m_\Sigma +
  2m'_\Sigma}$,
$a, b = 3, \cdots, 5$ and $i, j, k, l = 3, \cdots, 6$.
This superpotential implies a flat direction satisfying $\Lambda^{-5}
(B_6 \bar{B}^6 - \det M^a{}_{\! b}) + f \Phi = 0$.

Let us consider, among the vacua of Eq.(\ref{eff.pot}), the two vacua
which satisfy $\langle \Phi \rangle = 0$ or $\langle B_6 \rangle =
\langle \bar{B}^6 \rangle = 0$.

Vacuum (a): The vacuum with $\langle \Phi \rangle = 0$ is analyzed
in Ref.\cite{our2}.%
\footnote{Although the model in Ref.\cite{our2} does not contain the
  singlets $Tr \Sigma$ and $\Phi$, the vacuum considered in
  Ref.\cite{our2} is equivalent to that with $\langle \Phi \rangle =
  0$ in the present model.}
We see that $B_6$ and $\bar{B}^6$
have non-vanishing vacuum-expectation values leading to breaking of
the U(1)$_Y$ subgroup of SU$(5)_{GUT}$.
Thus we need to introduce an extra U$(1)_H$ gauge symmetry so as to
have the standard-model gauge group unbroken below the GUT scale.

Notice that this quantum vacuum is the same as the classical one,
which is consistent with the fact that the classical moduli space is
not altered by quantum corrections for the case of $N_f = N_c +1$
\cite{seiberg2} where $N_f$ and $N_c$ are the numbers of flavors and
colors of the massless hyperquarks, respectively.

Vacuum (b): The vacuum with $\langle B_6 \rangle = \langle
\bar{B}^6 \rangle = 0$ is analyzed in Ref.\cite{our3}.
That is
\begin{equation}
  \begin{array}{l}
    \label{quantumvacuum}
    \langle B_i \rangle = \langle \bar{B}^i \rangle = 0 \, ,\\
    \noalign{\vskip 1ex}
    \langle M^6{}_{\! a} \rangle = \langle M^a{}_{\! 6} \rangle =
    \langle M^6{}_{\! 6} \rangle = 0 \, ,\\
    \displaystyle \langle M^a{}_{\! b} \rangle = 
    {m_\Sigma \mu_\Sigma \over \lambda (m_\Sigma + 2m'_\Sigma)} \,
    \delta^a{}_{\! b} \, , \quad
    \langle \Phi \rangle = \frac{1}{f \Lambda^5}
    \left[ {m_\Sigma \mu_\Sigma \over \lambda
        (m_\Sigma + 2m'_\Sigma)}
    \right]^3 \, ,
  \end{array}
\end{equation}
where the GUT gauge group is broken down to the standard-model one,
namely SU(3)$_C \:\times$ SU(2)$_L \:\times$ U(1)$_Y$.
Thus, there is no need to introduce an extra U(1)$_H$, differently
from the previous phase (a).
An interesting point is that this quantum vacuum differs from the
classical one which satisfies $B_6 \bar{B}^6 - \det M^a{}_{\! b} = 0$.
This result agrees with the conclusion in Ref.\cite{seiberg2} for the
case of $N_f = N_c$.
Notice that the effective $N_f$ is three $(= N_c)$ in the present
phase since the vacuum-expectation value of $\Phi$ induces a mass for
$Q_\alpha^6$ and $\bar{Q}^\alpha_6$.

As noted in Ref.\cite{our3}, we have a pair of massless bound
states $B_6$ and $\bar{B}^6$ in this vacuum.
Since they have non-vanishing U(1)$_Y$ charges, they contribute to the 
renormalization-group equations of three gauge coupling constants in
the standard model.
A change of running of couplings threatens to destroy the GUT
unification of gauge coupling constants which is regarded as one
of the motivations for considering the SUSY-GUT as a unified theory.

However, it seems quite reasonable to assume that there are
nonrenormalizable operators in the superpotential suppressed by some
scale $M_0$ higher than the GUT scale (originating from gravitational
interactions, for example).
Among such operators we consider the lowest-dimensional
nonrenormalizable operator consistent with our gauge and global
symmetries which is to contain baryon superfields.
That is 
\begin{equation}
  \label{highdimop}
  \delta W = \frac{f'}{M^3_0} \epsilon^{\alpha \beta \gamma}
  \epsilon_{\alpha' \beta' \gamma'} (Q^I_\alpha Q^J_\beta Q^K_\gamma)
  (\bar{Q}^{\alpha'}_I \bar{Q}^{\beta'}_J \bar{Q}^{\gamma'}_K) \, .
\end{equation}
This interaction generates a mass term for $B_6$ and $\bar{B}^6$ in
the effective superpotential as
\begin{equation}
  \delta W_{eff} = \frac{f'}{M^3_0} B_6 \bar{B}^6 \, ,
\end{equation}
which corresponds to the physical mass for $B_6$ and $\bar{B}^6$
\begin{equation}
  \label{massB6}
  m_{B_6} \simeq \frac{f' \Lambda^4}{M^3_0} \, .
\end{equation}

If one takes $M_0$ in Eq.(\ref{massB6}) at the gravitational scale,
i.e. $M_0 \simeq 2 \times 10^{18}$ GeV, and $\Lambda \simeq 3 \times
10^{16}$ GeV, for example, one has the mass for $B_6$ and $\bar{B}^6$
$\sim 10^{11}$ GeV for $f' \sim {\cal O}(1)$.
This mass is too small compared with the GUT scale and the presence of
$B_6$ and $\bar{B}^6$ destroys the GUT unification of gauge coupling
constants.
However, since $M_0$ is given at the gravitational scale, it evolves
as the change of scale by renormalization effects.
Provided that the renormalized $M^R_0$ becomes about $10^{17}$ GeV at
the GUT scale,\,%
\footnote{Above the GUT scale the number of effectively massless
  flavors is six and the present model lies in the conformal window
  \cite{seiberg1}.
  Therefore, the model has a quasi-infrared fixed point.
  As pointed out in Ref.\cite{kogan} the renormalization factor $Z_Q$
  for the wave functions of quarks $Q$ and $\bar{Q}$ goes to vanish in
  the long distance.
  This suggests the renormalized mass $M^R_0 \equiv Z_Q^2 M_0$
  becomes smaller as the renormalization point is lowered.
  We also suspect that this kind of renormalization effects may be an
  origin of the GUT scale itself.}
we obtain  $m_{B_6} \sim 10^{15}$ GeV for $\Lambda \simeq 3 \times
10^{16}$ GeV and $f' \sim {\cal O}(1)$.
This result turns out to be consistent with the recent experimental
data on the three gauge coupling constants~\cite{our3}.

In both the phases (a) and (b), the colored Higgs $H_a$ and $\bar{H}^a$
acquire masses of the GUT scale with the composite states $M^6{}_{\! a}$
and $M^a{}_{\! 6}$ from the interactions in Eq.(\ref{eff.pot}), but the
Higgs doublets $H_I$ and $\bar{H}^I \, (I = 1, 2)$ remain massless
because of $\langle M^6{}_{\! 6} \rangle = 0$.

It is remarkable that the operator in Eq.(\ref{highdimop}) changes the 
quantum moduli space.
Actually, the vacuum (a) is no longer in the quantum moduli space,
while the vacuum (b) still remains there.
Since it is quite natural to consider that the operators suppressed by
the gravitational scale exist, there is a doubt as to the presence of
the vacuum (a).

\section{The dual model}
\label{sec:dual}

Nonabelian duality proposed in Ref.\cite{seiberg1} enables us to study 
supersymmetric nonabelian gauge theories with different color groups.
The duality means that an SU($N_c$) gauge theory with $N_f$ flavors of
quarks and an SU($N_f - N_c$) gauge theory with the same number of
quarks have the same behavior in the infrared limit, especially the
same moduli space of vacua.
Thus, one may investigate the dual theory instead of our original
model, which may even clarify the structure of the original model.
Furthermore, there is a possibility that we can throw a new light on
the origin of the fields in the original model, such as $H_I$ and
$\bar{H}^I$, as composite fields of the elementary hyperquarks in the
dual theory.

Now let us consider the nonabelian dual of the model in section
\ref{sec:original}, which is expected to coincide with the original
model as a low-energy effective theory.

The dual gauge group turns out to be SU(3), and thus we denote it as
SU$(3)_{\tilde H}$ in distinction with the original gauge group
SU(3)$_H$.
The dual model is described by a supersymmetric SU$(3)_{\tilde H}$
gauge theory with singlet chiral superfields $\sigma^A{}_{\! B}$ and
six flavors of dual hyperquarks $\bar{q}^\alpha_A$ and $q^A_\alpha$ in
the fundamental representations ${\bf 3}^*$ and {\bf 3}, respectively.
In addition we have to introduce Yukawa couplings between hyperquarks
and $\sigma^A{}_B$ to obtain a correct global symmetry and correct
vacua.

Then the dual superpotential to Eq.(\ref{orig.pot}) is given by
\begin{equation}
  \label{dual.pot}
  \begin{array}{rl}
    {\tilde W} &= {\tilde \lambda} \bar{q}^\alpha_A \sigma^A{}_{\!
      B} q^B_\alpha
    + \lambda \rho \Sigma^I{}_{\! J} \sigma^J{}_{\! I}
    + h \rho H_I \sigma^I{}_{\! 6}
    + h' \rho {\bar H}^I \sigma^6{}_{\! I} + f \rho \, \Phi
    \sigma^6{}_{\! 6} \\
    \noalign{\vskip 1ex}
    & \displaystyle + \frac{1}{2} m_\Sigma Tr(\Sigma^2) 
    + \frac{1}{2} m'_\Sigma(Tr \Sigma)^2 - \mu_\Sigma Tr\Sigma, \\
    \noalign{\vskip 0.5ex}
    & \hskip 5em (I, J = 1, \cdots , 5) ,
  \end{array}
\end{equation}
where $\rho$ denotes the duality scale to match the operator
dimensions in a correspondence~\cite{Intriligator}
\begin{equation}
  \label{duality}
  \rho \sigma^A{}_{\! B} \sim Q_\alpha^A \bar{Q}_B^\alpha \, , \ \
  (A, B = 1, \cdots 6).
\end{equation}
To obtain Eq.(\ref{dual.pot}) the hyperquarks $Q_\alpha^A$ and
$\bar{Q}_A^\alpha$ in Eq.(\ref{orig.pot}) are substituted with
$\sigma^A{}_{\! B}$ by means of the relation Eq.(\ref{duality}).

The fields in Eq.(\ref{dual.pot}) transform under a global U(1)$_A$
symmetry as
\begin{equation}
  \label{dualuonea}
  \begin{array}{ccl}
    \bar{q}^\alpha_I, q^I_\alpha & \rightarrow &
    \bar{q}^\alpha_I, q^I_\alpha , \\
    \noalign{\vskip 1ex}
    \bar{q}^\alpha_6, q^6_\alpha & \rightarrow &
    e^{-i \xi} \bar{q}^\alpha_6,
    e^{-i \xi} q^6_\alpha , \\ 
    \noalign{\vskip 1ex}
    \sigma^I{}_{\! J} & \rightarrow & \sigma^I{}_{\! J} , \\
    \noalign{\vskip 0.5ex}
    \sigma^I{}_{\! 6} , \sigma^6{}_{\! I} & \rightarrow & e^{i \xi}
    \sigma^I{}_{\! 6} , e^{i \xi} \sigma^6{}_{\! I}\\ 
    \noalign{\vskip 0.5ex}
    \sigma^6{}_{\! 6} & \rightarrow & e^{2 i \xi} \sigma^6{}_{\! 6} ,
  \end{array}
\end{equation}
with the transformation law in Eq.(\ref{uonea}).

Since the dual superpotential in Eq.(\ref{dual.pot}) looks
complicated, we reduce it by integrating out the superfields $H_I$,
${\bar H}^I$, $\Sigma$, $\Phi$, $\sigma^I{}_6$, $\sigma^6{}_I$ and
$\sigma^6{}_6$ to
\begin{equation}
  \label{dualpotential}
  {\tilde W}' =  {\tilde \lambda} \bar{q}^\alpha_I \sigma^I{}_{\!
    J} q^J_\alpha
  + \frac{1}{2} m_\sigma Tr(\sigma^2)
  + \frac{1}{2} m'_\sigma (Tr \sigma)^2 - \mu_\sigma
  Tr\sigma,
\end{equation}
where
\begin{equation}
  \label{dualdef}
  m_\sigma = - \frac{(\lambda \rho)^2}{m_\Sigma},  \quad
  m'_\sigma = \frac{(\lambda \rho)^2 m'_\Sigma}{(m_\Sigma + 5
    m'_\Sigma) \, m_\Sigma} , \quad
  \mu_\sigma = - \frac{\lambda \rho \mu_\Sigma}{m_\Sigma + 5 m'_\Sigma}.
\end{equation}
Notice that the sixth hyperquarks $\bar{q}^\alpha_6$ and $q^6_\alpha$
are massless.
If we integrate out $\sigma^I{}_{\! J}$, Eq.(\ref{dualpotential})
becomes a superpotential including only $\bar{q}^\alpha_I$ and
$q^I_\alpha$ with nonrenormalizable interactions $\displaystyle
\frac{1}{m_\sigma} (\bar{q}^\alpha_I q^J_\alpha)^2$ and $\displaystyle 
\frac{1}{m'_\sigma} (\bar{q}^\alpha_I q^J_\alpha)^2$.
From this viewpoint, the superfields $\sigma^I{}_{\! J}$ are
interpreted as composite states of $\bar{q}^\alpha_J$ and $q^I_\alpha$.

Let us study vacua corresponding to the original vacua given in the
previous section which satisfy
\begin{equation}
  \label{dualvacua}
  \displaystyle \langle \sigma^I{}_{\! J} \rangle =
  \frac{\mu_\sigma}{m_\sigma + 3 m'_\sigma}
  \left(
    \begin{array}{ccccc}
      1 & & & & \\
      & 1 & & & \\
      & & 1 & & \\
      & & & 0 & \\
      & & & & 0
    \end{array}
  \right) \, .
\end{equation}

The integration of massive hyperquarks $\bar{q}^\alpha_I$ and
$q^I_\alpha$ $(I = 1, 2, 3)$ leads to a low-energy effective theory.
Since this effective theory is an SU(3)$_{\tilde{H}}$ gauge theory with
three massless hyperquarks $\bar{q}^\alpha_A$ and $q^A_\alpha$ $(A =
4, 5, 6)$, its effective superpotential is obtained by means of
knowledge for the $N_f = N_c$ case in Ref.\cite{seiberg2}.
The effective superpotential is described by meson $M^i{}_{\! j}$,
baryon $B$ and antibaryon $\bar B$ chiral superfields,
\begin{eqnarray}
  \begin{array}{lll}
    & & M^i{}_{\! j} \sim q^i_\alpha
    \bar{q}^\alpha_j \, , \\
    \noalign{\vskip 1ex}
    & & \displaystyle B \sim \epsilon^{\alpha \beta \gamma}
    q^4_\alpha q^5_\beta q^6_\gamma \, ,
    \\
    \noalign{\vskip 1ex}
    & & \displaystyle \bar{B} \sim \epsilon_{\alpha \beta \gamma}
    \bar{q}_4^\alpha \bar{q}_5^\beta
    \bar{q}_6^\gamma \, , \\
    \noalign{\vskip 1ex}
  \end{array}
\end{eqnarray}
as follows:
\begin{equation}
  \label{dualEff.pot}
  \begin{array}{rl}
    {\tilde W}_{eff} = & X (B \bar{B} - \det M^i{}_{\! j} -
    \tilde{\Lambda}^6) 
    + \tilde{\lambda} \sigma^b{}_{\! a} M^a{}_{\! b} \\
    \noalign{\vskip 1ex}
    & \displaystyle + \frac{1}{2} \, m_\sigma
    \sigma^a{}_{\! b} \sigma^b{}_{\! a}
    + \frac{1}{2} \, 
    \frac{m_\sigma m'_\sigma}{m_\sigma + 3 m'_\sigma}
    ( \sigma^a{}_{\! a} )^2
    - \frac{m_\sigma \mu_\sigma}{m_\sigma + 3 m'_\sigma}
    \sigma^a{}_{\! a} \, ,
  \end{array}
\end{equation}
where $\tilde{\Lambda}$ denotes a dynamical scale of the low-energy
SU(3)$_{\tilde{H}}$ interactions, $a, b = 4, 5$, and $i, j = 4, 5,
6$.
This superpotential implies a flat direction satisfying $B \bar{B} -
\det M^i{}_{\! j} -\tilde{\Lambda}^6 = 0$.

We consider two vacua, $\langle M^6{}_6 \rangle= 0$ and $\langle B
\rangle = \langle \bar{B} \rangle = 0$, which have a pair of massless
Higgs doublets.
Here these Higgs doublets are all composite bound states of the dual
hyperquarks $M^6{}_{\! i} \sim q^6_\alpha \bar{q}^\alpha_i$ and
$M^i{}_{\! 6} \sim q^i_\alpha \bar{q}^\alpha_6 \ (i = 4, 5)$.

Vacuum (a'): The vacuum with $\langle M^6{}_6 \rangle = 0$
corresponds to the vacuum (a) in the previous section, where the
fields have the following vacuum-expectation values:
\begin{equation}
  \begin{array}{l}
    \langle B \rangle = \langle \bar{B} \rangle = \tilde{\Lambda}^3 \, 
        ,\\
    \displaystyle \langle M^a{}_{\! b} \rangle = 
    - \frac{m_\sigma \mu_\sigma}{\tilde{\lambda} (m_\sigma + 3
      m'_\sigma)} \, \delta^a{}_{\! b} \, , \\
    \noalign{\vskip 1ex}
    \langle M^6{}_{\! a} \rangle = \langle M^a{}_{\! 6} \rangle = 0, 
    \, \langle \sigma^a{}_{\! b} \rangle  = 0 \, .
  \end{array}
\end{equation}
In this vacuum the SU(5)$_{GUT}$ breaks down to SU(3)$_C \:\times$
SU(2)$_L$ since the baryon $B$ and the antibaryon $\bar{B}$ have
non-vanishing U(1)$_Y$ charges.
Therefore, we need an extra U(1)$_H$ as in the vacuum (a).
Interesting enough, although the point $\langle B \rangle = \langle
\bar{B} \rangle = 0$ and $\det M^i{}_{\! j} = 0$ with unbroken
U(1)$_Y$ is in the classical moduli space, this point disappears from
the moduli space by non-perturbative effects at the quantum level.
This phenomenon contrasts with the vacuum (b) in section
\ref{sec:original} in which the U(1)$_Y$ breaks down classically but
is restored quantum mechanically.

Although there is no problem phenomenologically if the U(1)$_H$ is
strong enough at the GUT scale, the U(1)$_H$ brings some theoretical
problems.
First of all, the U(1)$_H$ is not asymptotically free and its gauge
coupling constant blows up at some higher scale.\,%
\footnote{For $m_\sigma \sim 10^{18}$ GeV, the mass of the SU(2)$_L$ 
  triplet in $M^a{}_{\! b}$ is of the order of $\tilde{\Lambda}^2 /
  m_\sigma$ which may be smaller than the GUT scale.
  In this case, the unification of the three gauge coupling constants
  is realized within the experimental errors, even if the gauge
  coupling $g_H$ of U(1)$_H$ is not so large.
  Thus, it is possible that the coupling $g_H$ does not diverge below
  the Planck scale.}
Secondly, the charge quantization is left unexplained.

Vacuum (b'): The vacuum with $\langle B \rangle = \langle \bar{B}
\rangle = 0$ corresponds to the vacuum (b) in the original model.
We find the vacuum:
\begin{equation}
  \langle M^6{}_{\! 6} \rangle = - \frac{\tilde{\lambda}^2 (m_\sigma +
    3 m'_\sigma)^2}{m_\sigma^2 \mu_\sigma^2} ,
\end{equation}
and the other fields acquire the same expectation values as in the
vacuum (a') which breaks SU(5)$_{GUT}$ down to SU(3)$_C \:\times$
SU(2)$_L \:\times$ U(1)$_Y$.
Therefore, we do not need to introduce an extra U(1)$_H$.
As in (a') this quantum vacuum is different from the classical one
which satisfies $\det M^i{}_{\! j} = 0$.

As in the vacuum (b), there is a pair of massless baryon $B$ and
antibaryon $\bar{B}$ with non-vanishing U(1)$_Y$ charges.
Unlike in the original model, however, nonrenormalizable operators
generating the mass for $B$ and $\bar{B}$ are forbidden by the global
U(1)$_A$ symmetry in Eq.(\ref{dualuonea}).
The effective superpotential in Eq.(\ref{dualEff.pot}) says that
nonperturbative effects never generate the mass term of the baryon $B$ 
and the antibaryon $\bar{B}$ although the U(1)$_A$ is broken by
instanton effects.
Thus, there remains a pair of the baryons in the massless spectrum,
which renders the vacuum (b') unrealistic.
It also suggests that the short-distance behaviors of the original and
dual models are different.%
\footnote{If one adds a nonrenormalizable operator
  $\frac{\tilde{f'}}{M^3_0} q^4 q^5 q^6 \bar{q}_4 \bar{q}_5
  \bar{q}_6$, one reproduces the same low-energy physics as in the
  previous vacuum (b).
  However, short-distance physics are different from each other, since
  in the original model the global U(1)$_A$ is unbroken at the
  classical level whereas there is no such a symmetry in its dual
  model.}
On the other hand, any nonrenormalizable operators with the U(1)$_A$
symmetry do not affect the stability of the vacuum (a').
Indeed, it is rather guaranteed by an introduction of the term
$\tilde{\Phi} M^6{}_{\! 6}$ which leads to $\langle M^6{}_{\! 6} 
\rangle = 0$.

We note an interesting relation between the original and the dual
models.
In the original model the Higgs multiplets are regarded as elementary 
fields.
Whereas in the dual model the Higgs doublets (even including
$\sigma^I{}_{\! J}$) are composite states of the dual hyperquarks
$\bar{q}^\alpha_A$ and $q^A_\alpha$.
Moreover, all the Higgs multiplets, $H_I$, $\bar{H}^I$,
$\Sigma^I{}_{\! J}$ and $\Phi$, in the original model might be
regarded as composite states of the elementary hyperquarks in the dual
model.

To summarize, the dual model reproduces exactly the same low-energy
physics as the original model does.
This supports the correctness of the duality arguments.
We proceed to use this nonabelian duality as a powerful tool to
investigate supersymmetric gauge theories with other hypercolors in
the next section.

\section{Other hypercolors}
\label{sec:hypercolor}

In this section we consider supersymmetric SU($N_c$)$_H$ hypercolor
gauge theories other than the SU(3)$_H$.
We continue to restrict ourselves to the minimal case of six flavors
of hyperquarks.\,%
\footnote{We may consider more than six flavors.
  In fact, Ref.\cite{our1} deals with the case of seven flavors.
  Then the Peccei-Quinn symmetry is naturally accommodated, though the 
  low-energy spectrum is rather involved with two pairs of light Higgs 
  doublets.}

For the theories of SU($N_c$)$_H$ ($N_c \geq 5)$ we find that there
is no appropriate vacuum which breaks SU(5)$_{GUT}$ down to
standard-model gauge group by a similar argument to that in
Ref.\cite{our2, our3}.
Since the case of $N_c = 3$ is already analyzed in the previous
section, $N_c = 2$ and $4$ remain as possible gauge groups.

$(i)$ First we investigate an SU(4)$_H$ gauge theory.

Since this model is the same as the one in section \ref{sec:original}
except that the index $\alpha$ runs from $1$ to $4$, the
superpotential is written as Eq.(\ref{orig.pot}).
We consider a classical vacuum given by
\begin{equation}
  \begin{array}{l}
    \langle Q_\alpha^A \rangle = \left(
    \begin{array}{cccc}
      0 & 0 & 0 & 0 \\
      0 & 0 & 0 & 0 \\
      0 & v & 0 & 0 \\
      0 & 0 & v & 0 \\
      0 & 0 & 0 & v \\
      0 & 0 & 0 & 0 \\
    \end{array}
  \right),\
  \langle \bar{Q}^\alpha_A \rangle = \left(
  \begin{array}{cccccc}
    0 & 0 & 0 & 0 & 0 & 0 \\
    0 & 0 & v & 0 & 0 & 0 \\
    0 & 0 & 0 & v & 0 & 0 \\
    0 & 0 & 0 & 0 & v & 0
  \end{array}
\right), \\
\noalign{\vskip 2ex}
\displaystyle
\langle \Sigma^I{}_{\! J} \rangle
= \frac{\mu_\Sigma}{m_\Sigma + 2m'_\Sigma} 
\left(
  \begin{array}{ccccc}
    1 & & & & \\
    & 1 & & & \\
    & & 0 & & \\
    & & & 0 & \\
    & & & & 0
  \end{array}
\right), \\
\noalign{\vskip 1ex}
\langle H_I \rangle = \langle \bar{H}^I \rangle = 0 \, ,
\end{array}
\end{equation}
where 
$\displaystyle v = \sqrt{\frac{m_\Sigma \mu_\Sigma}{\lambda(m_\Sigma + 
    2m'_\Sigma)}}$ 
and $\Phi$ remains undetermined.
In this classical vacuum the gauge group is broken down as desired,
but this vacuum does not survive quantum corrections as we will see
below.

Since two hyperquarks $Q^I_\alpha$ and $\bar{Q}^\alpha_I$ $(I
= 1, 2)$ become massive, we can integrate them to obtain a low-energy
effective theory with $N_f = 4$.
The effective superpotential is described by gauge invariant operators
\begin{eqnarray}
  \begin{array}{lll}
    & & M^i{}_{\! j} \sim Q^i_\alpha \bar{Q}^\alpha_j \, , \\
    \noalign{\vskip 1ex}
    & & \displaystyle B \sim \epsilon^{\alpha \beta \gamma \delta}
    Q^3_\alpha Q^4_\beta Q^5_\gamma Q^6_\delta \, ,  \\
    \noalign{\vskip 1ex}
    & & \displaystyle \bar{B} \sim \epsilon_{\alpha \beta \gamma
      \delta} 
    \bar{Q}_3^\alpha \bar{Q}_4^\beta \bar{Q}_5^\gamma \bar{Q}_6^\delta 
    \, ,
  \end{array}
\end{eqnarray}
as follows:
\begin{eqnarray}
  \label{four.pot}
  \begin{array}{ll}
    W_{eff} = & X (B \bar{B} - \det M^i{}_{\! j} - \Lambda^6) +
    \lambda \Sigma^a{}_{\! b} M^b{}_{\! a} \\
    \noalign{\vskip 1ex}
    & + h H_a M^a{}_{\! 6} + h' \bar{H}^a M^6{}_{\! a}
    + f \Phi \, M^6{}_{\! 6} \\
    \noalign{\vskip 1ex}
    & \displaystyle - \frac{h h'}{m} (\bar{H}^1 H_1
    + \bar{H}^2 H_2) M^6{}_{\! 6} \\
    \noalign{\vskip 1ex}
    & \displaystyle + \frac{1}{2} \, m_\Sigma \Sigma^a{}_{\! b}
    \Sigma^b{}_{\! a}
    + \frac{1}{2} \, {m_\Sigma m'_\Sigma \over m_\Sigma + 2m'_\Sigma}
    ( \Sigma^a{}_{\! a} )^2
    - {m_\Sigma \mu_\Sigma \over m_\Sigma + 2m'_\Sigma}
    \Sigma^a{}_{\! a} \, ,
  \end{array}
\end{eqnarray}
where $\Lambda$ denotes a dynamical scale of the low-energy SU(4)$_H$
\ interactions, $\displaystyle m = \frac{\lambda \mu_\Sigma}{m_\Sigma
  +   2m'_\Sigma}$, $a, b = 3, 4, 5$ and $i, j = 3, \cdots, 6$.

From this effective superpotential in Eq.(\ref{four.pot}) we find a
quantum vacuum\,%
\footnote{If $\Phi$ were non-vanishing, the sixth hyperquarks
  $Q_\alpha^6$ and $\bar{Q}^\alpha_6$ would become massive.
  Then the vacuum becomes unstable quantum mechanically because the
  number of effective massless hyperquarks is three which is less than
  $N_c = 4$ \cite{affleck}.
  Therefore, $\Phi$ is fixed at the origin ($\langle \Phi \rangle =
  0$) in the stable vacuum.}
given by
\begin{eqnarray}
  \begin{array}{l}
    \langle X \rangle = 0 , \\
    \noalign{\vskip 0.5ex}
    \langle B \rangle = \langle \bar{B} \rangle = \Lambda^3 , \\
    \displaystyle \langle M^a{}_{\! b} \rangle = 
    - \frac{m_\Sigma \mu_\Sigma}{\lambda (m_\Sigma + 2m'_\Sigma)} \,
    \delta^a{}_{\! b} \, , \\
    \noalign{\vskip 0.5ex}
    \langle M^6{}_{\! a} \rangle = \langle M^a{}_{\! 6} \rangle =
    \langle M^6{}_{\! 6} \rangle = \langle \Sigma^a{}_{\! b} \rangle
    = 0 \, ,\\
    \noalign{\vskip 0.5ex}
    \langle H_a \rangle = \langle \bar{H}^a \rangle = \langle \Phi
    \rangle = 0 .
  \end{array}
\end{eqnarray}
Since $\langle M^6{}_{\! 6} \rangle$ is vanishing, we obtain exactly
massless Higgs doublets.
On the other hand, the baryon $B$ and the antibaryon $\bar{B}$ get
non-vanishing vacuum-expectation values and hence the U(1)$_Y$
subgroup of SU(5)$_{GUT}$ is broken down in this quantum vacuum in
contrast to the classical vacuum.

If one removes the term $f \Phi M^6{}_{\! 6}$ from
Eq.(\ref{four.pot}), there is a vacuum with $\langle M^6{}_{\! 6}
\rangle \neq 0$ and $\langle B \rangle = \langle \bar{B} \rangle  =
0$, that is,  U(1)$_Y$ is unbroken.
However, the Higgs doublets $H_I$ and $\bar{H}^I$ $(I = 1, 2)$
acquire masses from the interaction with $M^6{}_{\! 6}$ in this
vacuum, and hence there is no light Higgs doublet.

$(ii)$ Next we consider an SU(2)$_H$ gauge theory.

From the nonabelian duality, the SU(2)$_H$ gauge theory with six
flavors of hyperquarks is expected to be dual to the SU(4)$_H$ gauge
theory.

The superpotential in this theory is given by Eq.(\ref{dualpotential}) 
with $\alpha = 1, 2$.
The corresponding dual vacua satisfy
\begin{equation}
  \displaystyle \langle \sigma^I{}_{\! J} \rangle =
  \frac{\mu_\sigma}{m_\sigma + 3 m'_\sigma}
  \left(
    \begin{array}{ccccc}
      1 & & & & \\
      & 1 & & & \\
      & & 1 & & \\
      & & & 0 & \\
      & & & & 0
    \end{array}
  \right) \, .
\end{equation}
In these vacua three hyperquarks become massive and three remain
massless.
Integrating out the three massive hyperquarks $Q^I_\alpha$ and
$\bar{Q}^\alpha_I$ $(I = 1, 2, 3)$, we obtain the effective
superpotential written in terms of meson $M^i{}_{\! j}$, baryon $B_i$
and antibaryon $\bar{B}^i$ as\,%
\footnote{Their definitions are similar to those in
  Eq.(\ref{defMeson}) in the original SU(3)$_H$ model though there are
  no intrinsic difference between mesons and baryons in the SU(2)$_H$
  model.}
\begin{equation}
  \begin{array}{rl}
    {\tilde W}_{eff} =
    & \tilde{\Lambda}^{-5} ( B_i M^i{}_{\! j} \bar{B}^j - \det
    M^i{}_{\! j} )
    + \tilde{\lambda} \sigma_a{}^{\! b} M^a{}_{\! b} \\
    \noalign{\vskip 1ex}
    & \displaystyle + \frac{1}{2} \, m_\sigma
    \sigma^a{}_{\! b} \sigma^b{}_{\! a}
    + \frac{1}{2} \, 
    \frac{m_\sigma m'_\sigma}{m_\sigma + 3 m'_\sigma}
    ( \sigma^a{}_{\! a} )^2
    - \frac{m_\sigma \mu_\sigma}{m_\sigma + 3 m'_\sigma}
    \sigma^a{}_{\! a} \, ,
  \end{array}
\end{equation}
where $\tilde{\Lambda}$ denotes a dynamical scale of the low-energy
SU(2)$_H$ interactions, $a, b = 4, 5$, and $i, j = 4, 5, 6$.
By the same analysis as in section \ref{sec:original}, we find the
following quantum vacuum:
\begin{eqnarray}
  \begin{array}{l}
    \langle B_i \rangle = \langle \bar{B}^i \rangle =
    (
    \begin{array}{ccc}
      0 & 0 & v^2
    \end{array} 
    ) \, , \\
    \noalign{\vskip 0.5ex}
    \langle M^i{}_{\! j} \rangle = v^2 
    \left(
      \begin{array}{ccc}
        1 & & \\
        & 1 & \\
        & & 0
      \end{array}
    \right) \, , \\
    \noalign{\vskip 0.5ex}
    \langle \sigma^a{}_{\! b} \rangle = 0 \, ,\\
  \end{array}
\end{eqnarray}
where $v^2 = \displaystyle \frac{m_\sigma
  \mu_\sigma}{\tilde{\lambda}(m_\sigma + 3 m'_\sigma)}$.

Thus, we have a pair of massless Higgs doublets but the U(1)$_Y$
subgroup of SU(5)$_{GUT}$ is broken
by the non-zero expectation values of $B_6$ and $\bar{B}^6$.
This low-energy behavior of the theory is precisely the same as that
in the SU(4)$_H$ gauge theory as expected.

As for the vacuum stability, the perturbation such as $B_6 \bar{B}^6
\sim (q^4 q^5) (\bar{q}_4 \bar{q}_5)$ extinguishes the above vacuum.
Accordingly the SU(2)$_H$ model seems unrealistic.
On the other hand the corresponding vacuum in the SU(4)$_H$ model is
stable due to the global U(1)$_A$ symmetry.

\section{Conclusion}
\label{sec:conclusion}

In this paper we have investigated supersymmetric SU($N_c$)$_H$
hypercolor gauge theories with six flavors of hyperquarks.
There are two types of desirable vacua.
One type requires no extra U(1)$_H$ to have the standard-model gauge
group SU(3)$_C \:\times$ SU(2)$_L \:\times$ U(1)$_Y$ below the GUT
scale.
While the other needs an extra U(1)$_H$ gauge symmetry.
We have found that the SU(3)$_H$ gauge theory is a unique model for the
first type.
As for the second type of the vacua, SU(3)$_H$, SU(4)$_H$ and
SU(2)$_H$ are possible gauge groups.
We have also shown that some pairs of models are dual to each other
and thus they have the same low-energy physics.
It is intriguing that the Higgs superfields in one model are regarded
as the composite states of the elementary hyperquarks in another
model.

Even though each type is phenomenologically consistent, it has its own
shortcoming.
The first one needs nonrenormalizable interactions to remove unwanted
singlet baryons from the massless spectrum.
To derive the precise form of these interactions, it is necessary to
understand the physics at the Planck (gravitational) scale, yet still
unknown.
The second one requires an extra U(1)$_H$ gauge symmetry which seems
to have theoretical difficulties explained in the text.

One way to avoid the problem arising from the introduction of U(1)$_H$
is to embed U(1)$_H$ and SU(3)$_H$ into a larger simple group such as
SU(4)$_H$.
Here we show briefly an SU(4)$_H$ extension of the present SU(3)$_H
\:\times$ U(1)$_H$ model.

The model contains five flavors of hyperquark chiral superfields
$Q_{[\alpha \beta]}^I$ and $Q^{[\alpha \beta]}_I$ $(\alpha, \beta = 1,
\cdots ,4 ; I = 1, \cdots, 5)$ in the antisymmetric second rank tensor 
representation {\bf 6} of SU(4)$_H$ which transform as $\bf 5^*$ and
{\bf 5} under the GUT gauge group SU(5)$_{GUT}$, respectively.
We also introduce SU(5)$_{GUT}$ singlet chiral superfields $Q_{[\alpha
  \beta]}^6$ and $X^\alpha{}_{\! \beta}$ which transform as {\bf 6}
and an adjoint {\bf 15} of SU(4)$_H$, respectively.
We assume that the field $X^\alpha{}_{\! \beta}$ has a
vacuum-expectation value
\begin{equation}
  \langle X^\alpha{}_{\! \beta} \rangle \sim 
  \left (
    \begin{array}{cccc}
      1 & 0 & 0 & 0 \\
      0 & 1 & 0 & 0 \\
      0 & 0 & 1 & 0 \\
      0 & 0 & 0 & -3 \\
    \end{array}
  \right ) V_X
\end{equation}
at some scale $V_X$ above the GUT scale.
Then the hypercolor SU(4)$_H$ is broken down to SU(3)$_H \:\times$
U(1)$_H$ and the model includes the superfields in the original model
in section \ref{sec:original}.
Thus, we may have the consistent SU(3)$_H \:\times$ U(1)$_H$ model
discussed in this paper.
However, in addition to the GUT scale, we have to introduce a new
scale $V_X$.
An intriguing possibility is to identify the $V_X$ with the Planck
(gravitational) scale.
This could be done if the SU(3)$_H$ gauge coupling constant is closed 
to the infrared-stable fixed point.
However, it is beyond the scope of this paper to examine if it is
indeed the case.

We have always assumed the adjoint Higgs superfields $\Sigma^I{}_{\!
  J}$ of SU(5)$_{GUT}$ in this paper.
One of the purposes to introduce such fields is to eliminate unwanted
Nambu-Goldstone multiplets \cite{our1}.
We now comment that the Higgs fields $X^\alpha{}_{\! \beta}$ in the
adjoint representation of the hypercolor gauge group SU($N_c$)$_H$ may 
play the same role as $\Sigma^I{}_{\! J}$ \cite{hisano}.
A remarkable feature in this model is that if one imposes $N = 2$
extended supersymmetry, one has an SU(6) global symmetry.
The spontaneous breakdown of the global SU(6) naturally produces a
pair of massless Higgs doublets as Nambu-Goldstone multiplets
\cite{NGHiggs}.

So far we have discussed only the SU(5) gauge group as the GUT group.
We note, here, that our approach can also be applied to the SO(10)
GUT.
For example, we propose an SO(10)$_{GUT} \:\times$ SO(6)$_{H}$ gauge
theory with eleven hyperquarks $Q_\alpha^I$ $(I = 1, \cdots, 11)$ in the 
vector {\bf 6} representation of the SO(6)$_{H}$.
The first ten $Q_\alpha^I$ $(I = 1, \cdots, 10)$ transform as {\bf 10} 
under the SO(10)$_{GUT}$ and the last hyperquark $Q_\alpha^{11}$ is a
singlet of the SO(10)$_{GUT}$.
Instead of an adjoint Higgs $\Sigma^I{}_{\! J}$ of SO(10)$_{GUT}$, we
introduce a Higgs field $S_{IJ}$ in the symmetric second rank tensor
{\bf 54} and $S$ in the singlet {\bf 1} of the SO(10)$_{GUT}$.
It is very interesting that there is a vacuum in which the
SO(10)$_{GUT}$ breaks down to the Pati-Salam gauge group SU(4) %
$\times$ SU(2)$_L \:\times$ SU(2)$_R$ without having any unwanted
massless states except for the Higgs doublets transforming as ({\bf
  2}, {\bf 2}) under the SU(2)$_L \:\times$ SU(2)$_R$. 
The details of this model will be given in a forthcoming paper
\cite{our4}.

Finally, we should note that future experiments of the proton decay
might bring important informations to judge or distinguish the models
discussed in this paper.
The models without an extra U(1)$_H$ have the dangerous dimension-five 
operators \cite{dimfiveop} for nucleon decays and hence we have
unsuppressed proton decays \cite{our3}.
On the other hand, the models with an extra U(1)$_H$ have no such
operators and the proton decays are suppressed as explained in
Ref.\cite{our1, our2}.

\end{document}